\documentclass{article}
\usepackage{spconf,amsmath,amssymb,graphicx,courier,multirow,tikz,adjustbox,booktabs}
\usetikzlibrary{decorations.pathreplacing}
\usepackage{hyperref}
\usepackage{flushend}
\DeclareMathOperator*{\softmax}{softmax}
\DeclareMathOperator*{\argmin}{argmin}

\title{Overlap-aware low-latency online speaker diarization\\based on end-to-end local segmentation}
\name{
Juan M. Coria$^1$, Herv\'{e} Bredin$^2$, Sahar Ghannay$^1$, Sophie Rosset$^1$
\thanks{This work was granted access to the HPC resources of IDRIS under the allocation AD011012177 made by GENCI, and was partly funded by the French National Research Agency (ANR) through the PLUMCOT project (ANR-16-CE92-0025). Thanks to Antoine Laurent for running and sharing the VBx offline speaker diarization \textit{topline}.}
}
\address{
$^1$Universit\'{e} Paris-Saclay CNRS, LISN, Orsay, France \\
$^2$IRIT, Universit\'{e} de Toulouse, CNRS, Toulouse, France \\
$^1$\texttt{\small \{juan-manuel.coria, sahar.ghannay, sophie.rosset\}@lisn.upsaclay.fr} \\
$^2$\texttt{\small herve.bredin@irit.fr}
}
\begin{document}
%
\maketitle
\begin{abstract}
We propose to address online speaker diarization as a combination of incremental clustering and local diarization applied to a rolling buffer updated every 500ms. Every single step of the proposed pipeline is designed to take full advantage of the strong ability of a recently proposed end-to-end overlap-aware segmentation to detect and separate overlapping speakers. In particular, we propose a modified version of the statistics pooling layer (initially introduced in the x-vector architecture) to give less weight to frames where the segmentation model predicts simultaneous speakers. Furthermore, we derive \emph{cannot-link} constraints from the initial segmentation step to prevent two local speakers from being wrongfully merged during the incremental clustering step. Finally, we show how the latency of the proposed approach can be adjusted between 500ms and 5s to match the requirements of a particular use case, and we provide a systematic analysis of the influence of latency on the overall performance (on AMI, DIHARD and VoxConverse).
\end{abstract}
\section{Introduction}
\label{intro}

Speaker diarization aims at answering the question ``who spoke when'', effectively partitioning an audio sequence into segments with a particular speaker identity. Most dependable diarization approaches consist of a cascade of several steps~\cite{dia-review-2012, but-dihard1-clustering}: voice activity detection to discard \emph{non-speech} regions, speaker embedding~\cite{ivector, xvector} to obtain discriminative speaker representations, and clustering~\cite{but-dihard1-clustering, spectral, topline-dihard2} to group speech segments by speaker identity. The main limitation of this family of \emph{multi-stage} approaches relates to how they handle overlapped speech (which is known to be one of the main sources of errors): either they simply ignore the problem or they address it \emph{a posteriori} as a final post-processing step based on a dedicated overlapped speech detection module~\cite{otterson2007efficient, Bullock2020, horiguchi2021endtoend, segmentation-herve}. A new family of approaches have recently emerged, rethinking speaker diarization completely. Dubbed end-to-end diarization (EEND), the main idea of this approach is to train a single neural network -- in a permutation-invariant manner -- that ingests the audio recording and directly outputs the overlap-aware diarization output~\cite{pit, pit_attention}. We propose to meet half-way between \emph{multi-stage} and \emph{overlap-aware end-to-end} diarization and design a multi-stage pipeline where overlapped speech is a first-class citizen in every single step: from segmentation to incremental clustering. In particular, our first contribution (discussed in Section~\ref{sssec:embedding}) is a modified version of the statistics pooling layer (initially introduced in the x-vector architecture) to give less weight to frames where the intial segmentation step predicts simultaneous speakers.

Despite being competitive with \emph{multi-stage} approaches, the main limitation of the \emph{overlap-aware end-to-end} approaches is the strong assumption that the number of speakers is upper bounded or even known \emph{a priori}. While reasonable for some particular use cases (\emph{e.g.} one-to-one phone conversations), this assumption does not hold in many other situations (\emph{e.g.} physical meetings or conference calls). One solution to this problem is to augment \emph{end-to-end} approaches with mechanisms to automatically estimate the number of speakers. For instance, EEND-EDA~\cite{eda} extends EEND~\cite{pit, pit_attention} with a recurrent Encoder-Decoder network to generate a variable number of \emph{Attractors} -- similar to speaker centroids. \emph{Multi-stage} approaches usually do not suffer from this limitation as they rely on a clustering step for which a growing number of techniques exist to accurately estimate the number of speakers~\cite{park2019auto}. We propose to combine \emph{the best of both worlds}~\cite{end-to-end-plus-clustering} by first applying the end-to-end approach on audio chunks small enough to reasonably estimate an upper bound on the local number of speakers and, then only, apply global constrained clustering on top of the resulting local speakers. As discussed in Section~\ref{ssec:constrained_clustering}, we say that clustering is constrained because \emph{cannot-link} constraints are inferred from the output of the local end-to-end diarization. The main difference between this work and~\cite{end-to-end-plus-clustering} is that we target low-latency online speaker diarization while they address offline speaker diarization.

This work relies heavily on the speaker segmentation model introduced in~\cite{segmentation-herve} and summarized in Section~\ref{ssec:segmentation} for convenience. However, they address two very different problems with radically different constraints. While \cite{segmentation-herve} performs local offline speaker diarization of extremely short 5s chunks of audio, this work addresses online speaker diarization of (possibly infinite) audio streams. Hence, this work extends \cite{segmentation-herve} with a mechanism to track speakers over the duration of a conversation, with a latency much lower than 5s and real-time processing.

Low-latency online speaker diarization differs from its offline counterpart in several ways. While the latter assumes that the whole audio sequence is available at once (and hence can rely on multiple passes over the whole sequence to output its final prediction), the former ingests a possibly infinite audio stream and can only afford a short delay between when it receives a buffer of audio and when it outputs the corresponding prediction (without the option to correct it afterwards). These additional constraints prevent state-of-the-art \emph{multi-stage} approaches like VBx~\cite{vbx} from being used in that setting as they heavily rely on the possibility to pass several times over the audio sequence. 
EEND-like approaches are not suitable either because they expect large chunks of audio (30 seconds or more), leading to prohibitively high latency. One notable exception is FlexSTB~\cite{stb2} that astutely relies on an adaptive internal buffer to both simulate large audio chunks and support low (1s) latency. 

A comprehensive set of experiments on AMI, DIHARD~II, DIHARD~III and VoxConverse datasets is reported and discussed in Section~\ref{sec:results} -- where FlexSTB and a state-of-the-art offline approach based on VBx~\cite{vbx} respectively serve as baseline and \emph{topline}. In particular, we show how the latency of the proposed approach can easily be adjusted (without retraining) between 500ms and 5s to match the requirements of a particular use case.

\section{Overlap-aware online diarization}

\begin{figure}[htb]
    \centering
    \includegraphics[trim=0.0cm 0.0cm 0.0cm 0.0cm,clip,width=0.8\linewidth]{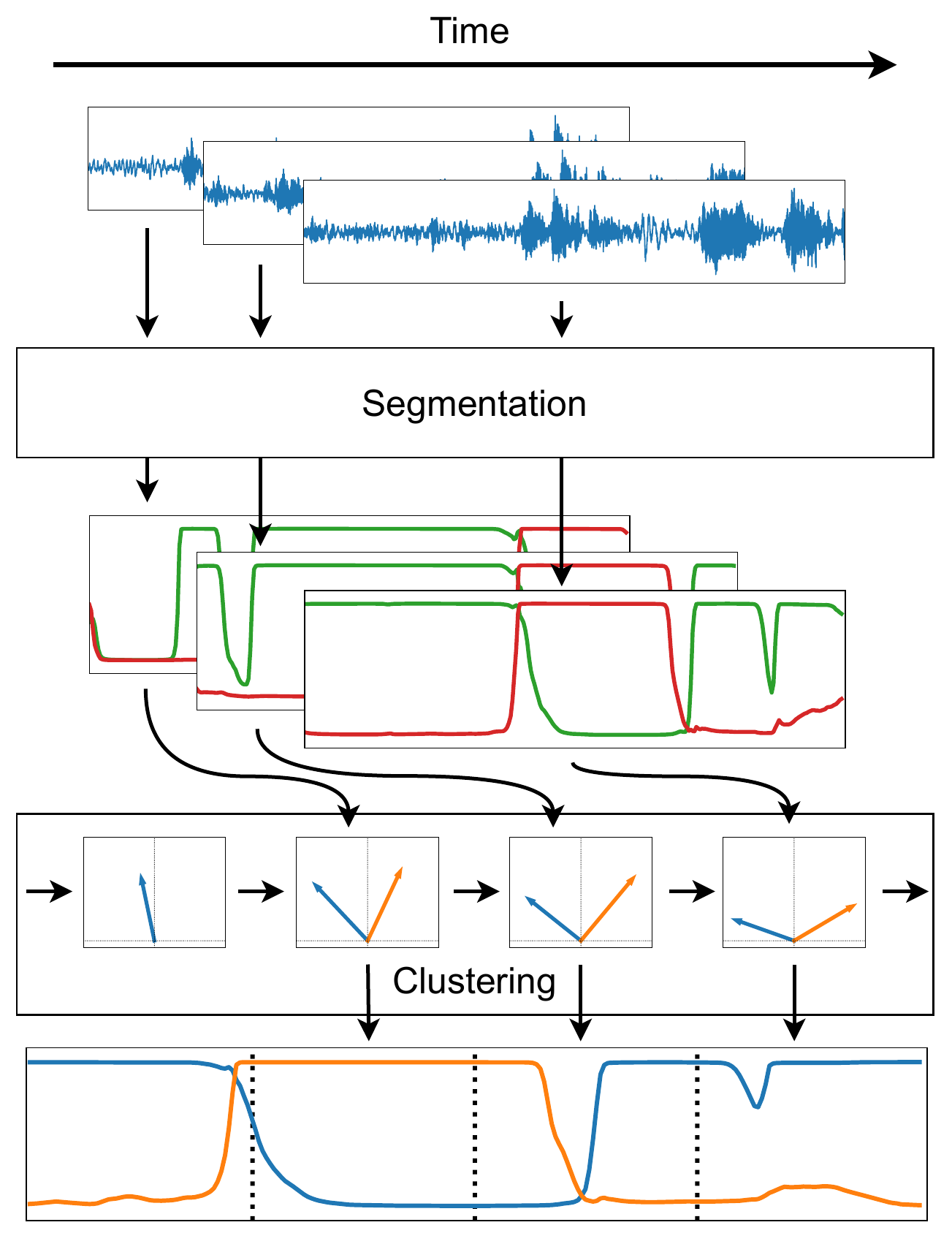}
    \caption{Block diagram of the proposed approach.}
    \label{fig:blocks}
\end{figure}

As depicted in Figure~\ref{fig:blocks}, we propose to address online speaker diarization as the iterative interplay between two main steps: segmentation and incremental clustering. Every few hundred milliseconds (500ms in our case), the segmentation module first performs a fine-grained overlap-aware diarization of a 5s rolling buffer. This local diarization is then ingested by the incremental clustering module that relies on speaker embeddings to map local speakers to the appropriate global speakers (or create new ones), before updating its own internal state.

\subsection{Segmentation}
\label{ssec:segmentation}

The segmentation step is the direct application of the end-to-end speaker segmentation neural network introduced in~\cite{segmentation-herve}, used to obtain a fine-grained local speaker diarization. 
As depicted in Figure~\ref{fig:blocks}, it ingests the 5s audio rolling buffer and outputs speaker activity probabilities $\{ \mathbf{s}_1, \ldots, \mathbf{s}_F \}$ where $F$ is the number of output frames and $\mathbf{s}_f \in [0,1]^{K_\text{max}}$, with $K_\text{max}$ the estimated maximum number of different speakers that a 5s chunk may contain ($K_\text{max}= 4$ in our case). Speakers whose activity probability exceeds a tunable threshold $\tau_{\mathrm{active}}$ at least once during the chunk constitute the set of local speakers. Inactive speakers are simply discarded.

Active speaker probabilities are then passed unchanged (i.e. with continuous values between 0 and 1) to the incremental clustering step. 
In particular, it means that overlapping speech (i.e. when two or more speakers have high probabilities simultaneously) is handled from the very beginning of the pipeline. This is in contrast with most dependable speaker diarization approaches that handle overlapping speech as a post-processing step~\cite{topline-dihard2, segmentation-herve}. This early detection of overlapping speech will prove very useful for the incremental clustering. 

\subsection{Incremental clustering}
\label{sec:clustering}

\begin{figure}[htb]
    \centering
    \includegraphics[trim=0.0cm 1.4cm 0.0cm 0.25cm,clip,width=\linewidth]{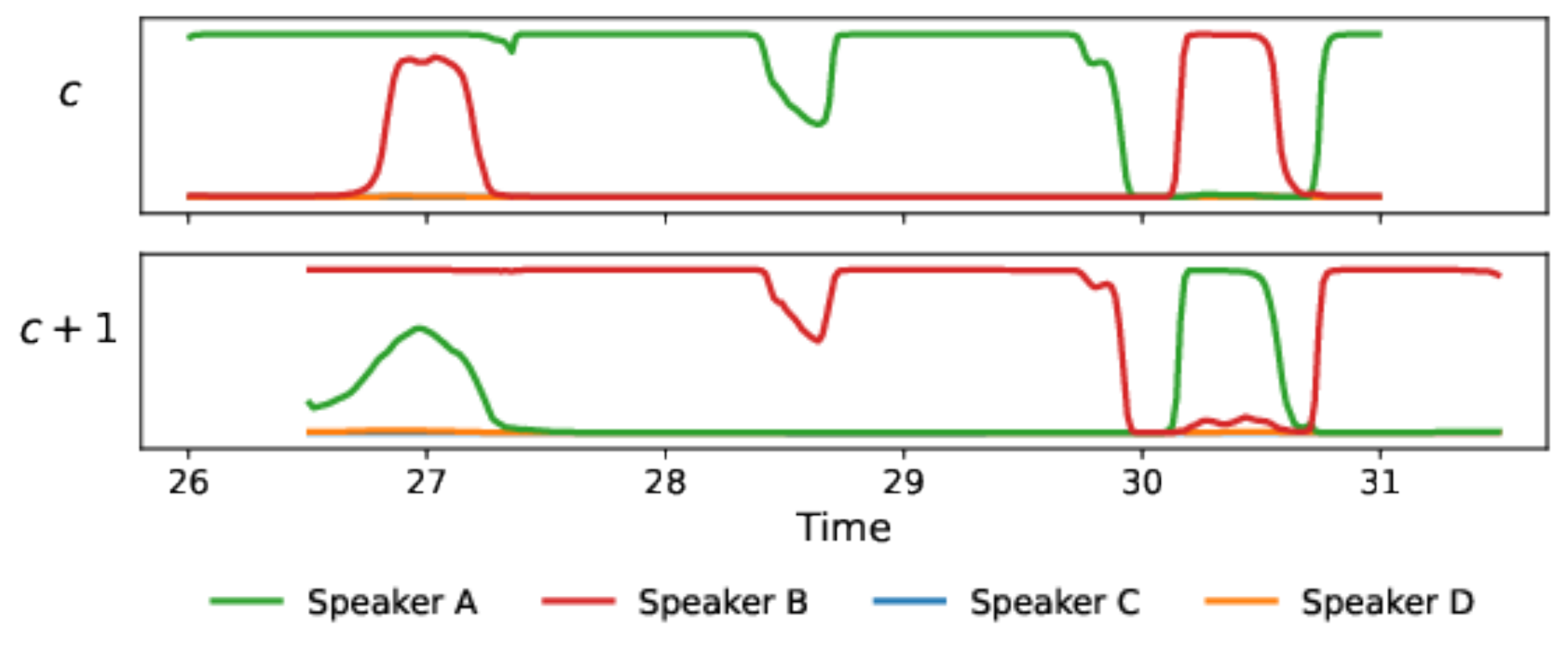}
    \caption{Actual output $\mathbf{s}_f$ of the segmentation model on two consecutive positions of the 5s rolling buffer in file \texttt{\small DH\_DEV\_0001} from DIHARD~III. Because of permutation-invariant training, green and red speakers are swapped.}
    \label{fig:swapped}
\end{figure}

Because the segmentation model is trained in a permutation-invariant manner and applied locally to the rolling buffer, one cannot guarantee that one particular speaker consistently activates the same index over time. 
Figure~\ref{fig:swapped} illustrates this limitation for two states of the rolling buffer: despite being only 500ms apart from each other and therefore having most of their audio content in common, notice how both active speakers are swapped. This section describes how we use incremental clustering to circumvent this limitation by tracking speakers (and detecting new ones) over the whole duration of the audio stream. 

\subsubsection{Segmentation-driven speaker embedding}
\label{sssec:embedding}

Like most recent speaker diarization systems, we rely on neural speaker embeddings to represent and compare speakers.
Our model is based on the canonical x-vector TDNN-based architecture, with the difference that the statistics pooling layer~\cite{xvector} is modified to return the concatenation of weighted mean $\mu_k$ and weighted standard deviation $\sigma_k$ for each active speaker $k$ -- instead of the regular mean $\mu$ and standard deviation $\sigma$:
\begin{eqnarray}
    \mu = \frac{\sum_f \mathbf{x}_f}{F}  & \longrightarrow & \mu_k = \frac{\sum_f w_{fk} \cdot \mathbf{x}_f}{\sum_f w_{fk}} \\
    \sigma^2 = \frac{\sum_f \left(\mathbf{x}_f - \mu\right)^2}{F - 1} & \longrightarrow & \sigma_k^2 = \frac{\sum_f w_{fk} \cdot \left(\mathbf{x}_f - \mu_k\right)^2}{\left(\sum_f w_{fk}\right) - \frac{\sum_f {w_{fk}}^2}{\sum_f w_{fk} }} \nonumber
\end{eqnarray}

\noindent where $\mathbf{x}_f$ is the output of frame $f$ of the last TDNN layer. One straightforward option is to derive $w_{fk}$ from the speaker activity probability and use $w_{fk} = s_{fk}$ directly, so that the final (pooled) speaker embedding mostly relies on frames where the segmentation model is confident that speaker $k$ is active. This generates exactly one embedding per active speaker in the current buffer, even when split into multiple speech turns (\textit{e.g.} the red speaker in the lower row of Figure~\ref{fig:swapped}).

\begin{figure}[htb]
    \centering
    \includegraphics[trim=0.0cm 1.4cm 0.0cm 0.25cm,clip,width=\linewidth]{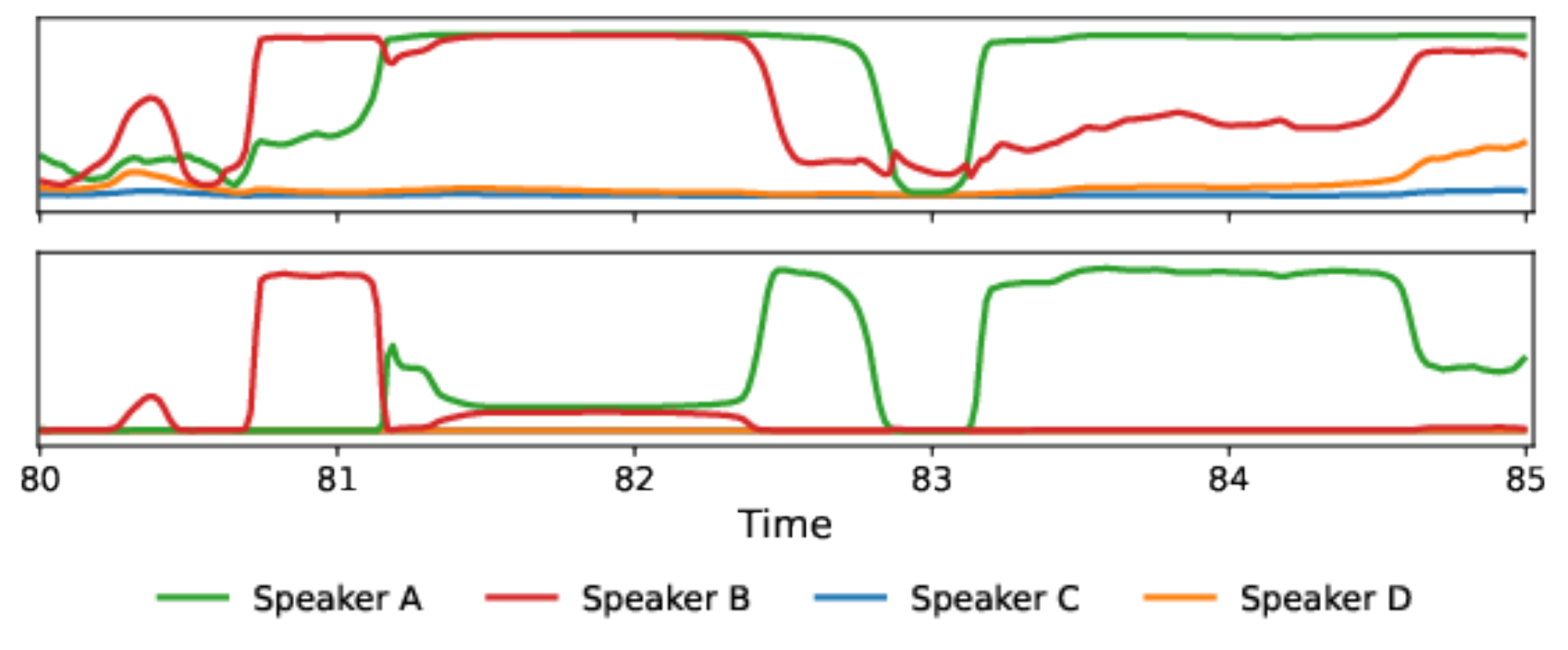}
    \caption{Effect of Equation~\ref{eq:overlap} with $\gamma = 3$ and $\beta = 10$. Top row is the actual output $\mathbf{s}_f$ of the segmentation model on a 5s excerpt of file \texttt{\small DH\_DEV\_0007} from DIHARD~III. Bottom row depicts weights $\mathbf{w}_f$ used for statistics pooling. Both low confidence and overlapped speech regions are weighed down.}
    \label{fig:overlap}
\end{figure}

Furthermore, as summarized in~\cite{segmentation-herve}, the segmentation model is also very good at detecting overlapped speech regions (where two or more speakers are active simultaneously). Therefore, another option is to make the speaker embedding focus on frames where it is confident that speaker $k$ is the only active speaker:

\begin{equation}
\label{eq:overlap}
    \mathbf{w}_f = \left({\mathbf{s}_f} \cdot \softmax_k \left(\beta \cdot \mathbf{s}_f\right)\right)^\gamma
\end{equation}

\noindent where the effect of this transformation is illustrated in Figure~\ref{fig:overlap}. The use of $\softmax_k$ weighs down frames where two or more speakers are active, and the exponent $\gamma > 1$ weighs down frames where the segmentation model is not quite confident about the activity of a speaker. Embeddings extracted with this weighing scheme are called \textit{overlap-aware speaker embeddings} in the rest of the paper.

\subsubsection{Constrained incremental clustering}
\label{ssec:constrained_clustering}

Given the initial content of the rolling buffer, the segmentation and embedding steps are combined to extract one embedding for each active speaker in the first 5s of the audio stream. These speaker embeddings $\{c_1, \ldots, c_K\}$ are stacked to form the initial centroid matrix $C$ with shape $K \times D$ where $K$ is the number of active speakers so far, and $D$ is the dimension of the speaker embedding.

Every few hundred milliseconds (e.g. 500ms), the rolling buffer is updated, and the segmentation and speaker embedding steps are combined to extract one embedding for each of the $K_\text{buffer} \leq K_\text{max}$ locally active speakers. Those $K_\text{buffer}$ speaker embeddings are then compared to the current state of the centroid matrix $C$ to find the optimal mapping $m^*$ between local and global speakers. Denoting $d\left(c, e\right)$ the distance between centroid $c$ and local speaker embedding $e$, one option is to assign the $k$th local speaker to the closest centroid:
\begin{equation}
\label{eq:naive-map}
m^*(k) = \argmin_{c \in C} d(c, e_k)
\end{equation}

Yet, this simple option does not take full advantage of the output of the segmentation model, as two local speakers might end up being assigned to the same centroid. This would be in contradiction to the output of the segmentation model that already chose to discriminate local speakers. Therefore, we add the constraint that any two local speakers cannot be assigned to the same centroid, while keeping the objective of minimizing the overall distance between local speakers and their assigned centroids:
\begin{equation}
    m^* = \argmin_{m \in \mathcal{M}} \sum_{k} d(m(k), e_k)
\end{equation}
\noindent where $\mathcal{M}$ is the set of mapping functions between local speakers and centroids with the following property:
\begin{equation}
    k \neq k' \implies m(k) \neq m(k')
\end{equation}
\noindent In practice, this optimal mapping is obtained by applying the Hungarian algorithm on the speaker-to-centroid distance matrix, and can be seen as an incremental clustering step with \textit{cannot-link} constraints.

\subsubsection{Detecting new speakers and updating centroids}

Once the optimal mapping $m^*$ is determined, for any given local speaker $k$ and their local embedding $e_k$
\begin{itemize}
    \item if $d(m^*(k), e_k) > \delta_{\mathrm{new}}$, they are marked as \textit{new speaker} (i.e. it is the first time they are active since the beginning of the audio stream) and their embedding~$e_k$ is appended to the pool of centroids: $$C \leftarrow C \cup e_k$$
    \item otherwise, they are marked as \textit{returning speaker}, and their embedding~$e_k$ is used to update the corresponding centroid. 
\end{itemize}

Because of the weighing scheme described in Section~\ref{sssec:embedding}, the quality of a speaker embedding $e_k$ is expected to be positively correlated with the estimated duration during which local speaker $k$ is active: $\Delta_k = \sum_f s_{fk}$. Therefore, we propose to only update a centroid when this duration is long enough:
\begin{equation}
c_{m^*(k)} \leftarrow \begin{cases}
             \displaystyle c_{m^*(k)} + e_k  & \text{ if } \displaystyle \Delta_k > \rho_{\mathrm{update}} \\ 
             \displaystyle c_{m^*(k)} & \text{ otherwise}
\end{cases}
\label{eq:centroid_update}
\end{equation}
where $\rho_\mathrm{update}$ is the minimum duration below which a speaker embedding $e_k$ is considered to be too noisy to help refine the centroid. Equation~\ref{eq:centroid_update} assumes that speaker embeddings $e_k$ are unit-normalized and optimized for cosine similarity.

\subsection{Adjusting the latency}
\label{ssec:latency}

Even though the whole $[t - 5s, t]$ buffer is used to extract embeddings and assign local speakers to an existing (or new) cluster, only the (active) speaker activity probabilities $s_{fk}$ at its rightmost part $[t - \lambda, t]$ are 
output: $\lambda$ effectively controls the latency of the whole system.

The lowest possible value for $\lambda$ corresponds to the period between two consecutive updates of the rolling buffer (500ms in our case). In this configuration, the rightmost parts of two consecutive buffer states $[t - 5s, t]$ and $[t + \lambda - 5s, t + \lambda]$ do not overlap: $[t - \lambda, t]$ and $[t, t + \lambda]$. Therefore, they are simply concatenated and frame-level speaker activity probabilities are passed through a final thresholding step. Local speaker $k$ is marked as active at frame $f$ if $s_{fk} > \tau_{\mathrm{active}}$. 

The careful reader might have noticed that, at the very beginning of the audio stream, the initial buffer must be filled entirely before a first output can be provided -- effectively leading to a much larger latency of 5s, an order of magnitude larger than the promised $\lambda=500$ms. However, once this initial $5s$ warm-up period has passed, the latency is indeed $\lambda=500$ms. If having a low latency from the very beginning of the stream is critical, one can simply left-pad the $[0, \lambda]$ initial incomplete buffer with zeros. 

\begin{figure}[htb]
    \centering
    \includegraphics[trim=0.0cm 0.4cm 0.0cm 0.25cm,clip,width=\linewidth]{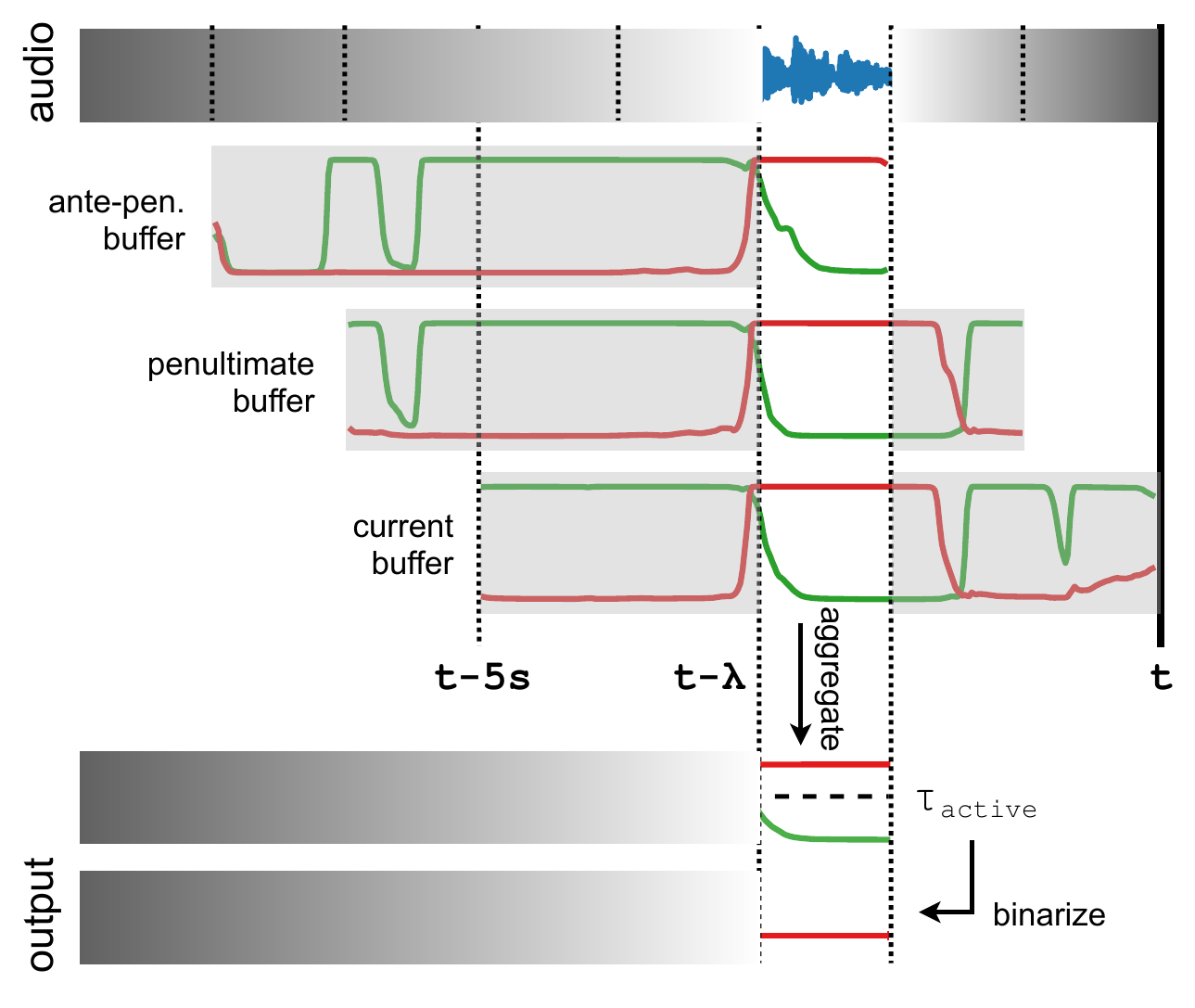}
    \caption{Depending on the allowed latency~$\lambda$ (between 500ms and 5s), multiple positions of the rolling buffer can be aggregated to obtain a (hopefully better) prediction. Speakers are colored according to the cluster they were found to belong to.}
    \label{fig:aggregation}
\end{figure}

Figure~\ref{fig:aggregation} shows that, for cases where longer latency $\lambda$ is permitted, several positions of the rolling buffer can be combined in an ensemble-like manner to obtain a more robust output. In practice, for a given frame $f$, the final speaker activity probabilities are computed as the average of the speaker activity probabilities obtained from each buffer position.

\section{Experiments}

\begin{table*}[t!]
  \centering
  \begin{adjustbox}{width=\textwidth}
  \begin{tabular}{l|c|ccc|c|ccc|c|ccc|c|ccc|c}
    \toprule
    \multirow{2}{*}{\textbf{System}} & \multirow{2}{*}{\textbf{Latency}} & \multicolumn{4}{c}{\textbf{DIHARD III} \cite{dihard3}} & \multicolumn{4}{|c}{\textbf{AMI} \cite{ami, vbx}} & \multicolumn{4}{|c}{\textbf{VoxConverse} \cite{voxconverse}} & \multicolumn{4}{|c}{\textbf{DIHARD II} \cite{dihard2}} \\ 
    & & FA & Miss. & Conf. & DER & FA & Miss. & Conf. & DER & FA & Miss. & Conf. & DER & FA & Miss. & Conf. & DER \\
    \midrule
    VBx~\cite{vbx} & $\infty$ & 3.6 & 12.5 & 6.2 & 22.3 & 3.1 & 17.2 & 3.8 & 24.1 & 3.1 & 4.6 & 3.4 & 11.1 & 5.0 & 15.3 & 7.4 & 27.7 \\
    \; $\hookrightarrow$ w/ overlap-aware segmentation~\cite{segmentation-herve} & $\infty$ & 4.7 & 9.7 & 4.9 & \textbf{19.3} & 4.3 & 10.9 & 4.7 & \textbf{19.9} & 4.6 & 3.0 & 3.5 & \textbf{11.1} & 5.6 & 13.5 & 7.1 & \textbf{26.3} \\
    \midrule
    \textbf{Ours} & 5s & 5.3 & 10.0 & 9.7 & \textbf{25.0} & 5.0 & 10.0 & 12.4 & \textbf{27.5} & 3.8	& 4.9 &8.2 & \textbf{16.8} & 5.7 & 14.0 & 14.4 & \textbf{34.1} \\
    \; $\hookrightarrow$ w/o overlap-aware embedding & 5s & 4.6 & 11.3 & 9.3 & 25.3 & 3.0 & 16.0 & 11.6 & 30.5 & 4.1 & 5.1 & 11.2 & 20.4 & 5.1 & 15.5 & 13.6 & 34.3 \\
    \; $\hookrightarrow$ w/ oracle segmentation & 5s & 2.1 & 1.4 & 6.9 & 10.4 & 1.0 & 1.1 & 15.5 & 17.7 & 0.5 & 0.7 & 9.1 & 10.3 & 2.2 & 1.6 & 12.0 & 15.8 \\
    \midrule
    FlexSTB~\cite{stb2} & 1s & \tiny{NA} & \tiny{NA} & \tiny{NA} & \tiny{NA} & \tiny{NA} & \tiny{NA} & \tiny{NA} & \tiny{NA} & \tiny{NA} & \tiny{NA} & \tiny{NA} & \tiny{NA} & \tiny{NA} & \tiny{NA} & \tiny{NA} & 36.0 \\
    \textbf{Ours} & 1s & 6.2 & 9.7 & 11.8 & 27.6 & 6.6 & 9.4 & 14.4 & 30.4 & 5.1 &3.3 & 11.7 & 20.1 & 5.8$^{\bullet}$ & 14.4$^{\bullet}$ & 14.9$^{\bullet}$ & \textbf{35.1}$^{\bullet}$ \\
    \bottomrule
  \end{tabular}
  \end{adjustbox}
  \caption{Experimental results on test sets. FA, Miss. and Conf. stand for false alarm, missed detection and speaker confusion rates respectively. ($^\bullet$ = hyper-parameters optimized with latency $\lambda = 1$s for fair comparison)}
  \label{tab:results}
\end{table*}

\subsection{Datasets}
\label{ssec:datasets}

We ran experiments on three different datasets covering a wide range of domains and number of speakers.\\

\noindent\textit{DIHARD~III}~\cite{ryant2020,DIHARD3_Evaluation_Plan} does not provide a \textit{training} set. Therefore, we split its \textit{development} set into two parts: 192 files used as \textit{training} set, and the remaining 62 files used as a smaller \textit{development} set. The latter is simply referred to as \textit{development} sets in the rest of the paper. When defining this split (shared at {\footnotesize{\texttt{\href{https://huggingface.co/pyannote/segmentation}{huggingface.co/pyannote/segmentation}}}}), we made sure that the 11 domains were equally distributed between both subsets. The \textit{test} set is kept unchanged. We also report performance on \textit{DIHARD~II}~\cite{dihard2} for comparison with FlexSTB~\cite{stb2}.\\

\noindent\textit{VoxConverse} does not provide a proper \textit{training} set either~\cite{voxconverse}. Therefore, we also split its \textit{development} set into two parts: the first 144 files ({\scriptsize\texttt{abjxc}} to {\scriptsize\texttt{qouur}}, in alphabetical order) constitute the \textit{training} set, leaving the remaining 72 files ({\scriptsize\texttt{qppll}} to {\scriptsize\texttt{zyffh}}) for the actual \textit{development} set. Furthermore, multiple versions of VoxConverse \emph{test} set have been circulating: we rely on version \texttt{0.0.2} available at {\footnotesize{\texttt{\href{https://github.com/joonson/voxconverse}{github.com/joonson/voxconverse}}}}.\\

\noindent \textit{AMI} provides an official \{\textit{training}, \textit{development}, \textit{test}\} partition of the Mix-Headset audio files~\cite{ami} that is further described as the \emph{Full} partitioning in~\cite{vbx}.

\subsection{Implementation details}

We use the pretrained segmentation model available at {\footnotesize{\texttt{\href{https://hf.co/pyannote/segmentation}{hf.co/pyannote/segmentation}}}}, which was trained on the composite training set made of the union of AMI, DIHARD~III, and VoxConverse respective training sets. It ingests 5~second audio chunks and outputs one prediction every 16ms with $K_\text{max} = 4$ speakers. More details about the training process can be found in~\cite{segmentation-herve}. 
The speaker embedding model is based on the canonical x-vector TDNN-based architecture~\cite{xvector}, but with filter banks replaced by trainable SincNet features~\cite{sincnet}. It was trained with additive angular margin loss~\cite{aam} using chunks of variable duration (from 2 to 5 seconds) drawn from VoxCeleb~\cite{voxceleb1,voxceleb2}, augmented with reverberation based on impulse responses from \textit{EchoThief} and~\cite{TraerE7856}, and additive background noise from MUSAN~\cite{musan}. It reaches an equal error rate of $2.8\%$ on VoxCeleb 1 test set using cosine distance only. We share the pretrained model and more details about the training process at {\footnotesize{\texttt{\href{https://hf.co/pyannote/embedding}{hf.co/pyannote/embedding}}}}.
Weights used in the statistics pooling layer of the \textit{overlap-aware speaker embeddings} were obtained with $\gamma = 3$ and $\beta = 10$. Those values were not optimized with the rest of the hyper-parameters. Instead, we handpicked them based on examples like the one in Figure~\ref{fig:overlap}.

\subsection{Experimental protocol}

While the same pretrained segmentation and embedding models were used for all three datasets, we rely on their respective development sets to optimize hyper-parameters ($\tau_{\mathrm{active}}$, $\rho_{\mathrm{update}}$ and $\delta_{\mathrm{new}}$) specifically for each dataset. More precisely, we use the {\footnotesize{\texttt{\href{https://github.com/pyannote/pyannote-pipeline}{pyannote.pipeline}}}} optimization toolkit that relies on a tree-structured Parzen estimator algorithm~\cite{tpe} to minimize the overall diarization error rate (DER) -- computed with {\footnotesize{\texttt{\href{https://pyannote.github.io/pyannote-metrics}{pyannote.metrics}}}}~\cite{pyannote.metrics} without any forgiveness collar and including overlapped speech regions. 
To ensure a fair comparison between different approaches, the  optimization process is applied for all of them independently. In other words, it means that every \textit{row} $\times$ \textit{dataset} entry in Table~\ref{tab:results} results from one dedicated optimization process.
This includes the offline \emph{topline}, the proposed online approach and its ablative variants, but excludes both FlexSTB (as we unfortunately did not have access to its implementation) and experiments on DIHARD~II (where we use respective hyper-parameters tuned for DIHARD~III).

\begin{figure*}[t!]
    \centering
    \includegraphics[trim=0.0cm 0.5cm 0.0cm 0.25cm,clip,width=\textwidth]{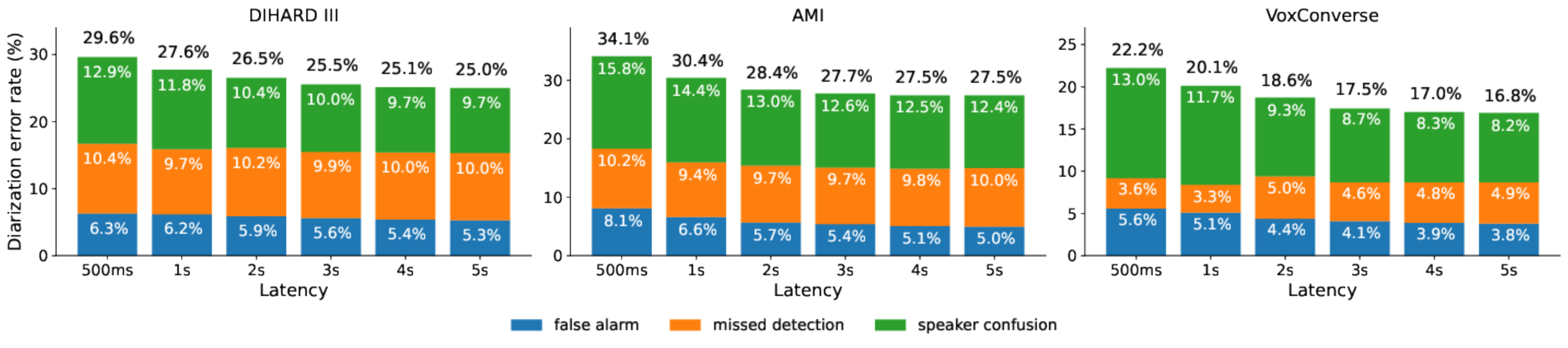}
    \caption{Impact of latency on the overall performance.}
    \label{fig:latency}
    \vspace{-0.3cm}
\end{figure*}

\vspace{-0.2cm}
\section{Results and discussion}
\label{sec:results}

\vspace{-0.2cm}
Table~\ref{tab:results} summarizes the whole set of experiments. \\

\vspace{-0.2cm}
\noindent \textbf{Offline vs. online.} We start by reporting the performance of a strong offline \textit{topline} that consists of VBx~\cite{vbx} followed by the overlap-aware resegmentation step introduced in~\cite{segmentation-herve}. Because this latter resegmentation step relies on the exact same pretrained segmentation model as our proposed approach, most of the reported  decrease in performance with $\lambda=5$s is caused by speaker confusion errors (relative $+100\%$ for DIHARD~III, $+160\%$ for AMI, $+130\%$ for VoxConverse). Incremental clustering still has a long way to go to be on par with offline multi-pass clustering. \\

\vspace{-0.2cm}
\noindent \textbf{Overlap-aware speaker embedding}. The first ablative experiment shows that the overlap-aware weighing scheme introduced in Equation~\ref{eq:overlap} brings a relative performance improvement of 10\% on AMI, 18\% on VoxConverse and 1\% on DIHARD~III. Given that they respectively contain 17\%, 3\%, and 11\% of overlapped speech, there is still room for improvement on this particular aspect. In particular, while we handcrafted this weighing scheme, it should be possible to train the segmentation and speaker embedding models jointly for the latter to fully take advantage of the former's capability at detecting and separating simultaneous speakers.\\

\vspace{-0.3cm}
\noindent \textbf{Overlap-aware speaker segmentation}. In a second ablative experiment, we replace the segmentation model by an oracle that provides perfect binary (\emph{i.e.} $s_{fk} \in \{0, 1\}$) overlap-aware segmentation. As expected, missed detection is where most of the difference occurs (caused by overlapped speech), while speaker confusion only marginally improves. The community has yet to solve the problem of overlapped speech detection. \\

\vspace{-0.3cm}
\noindent \textbf{Adjustable latency}. Figure~\ref{fig:latency} shows how the performance of our online approach evolves as we decrease the allowed latency from $\lambda=5$s to $\lambda=500$ms.
Speaker confusion error rate consistently increases as the latency decreases -- while false alarm and missed detection remain constant. This can be explained by the ensemble-like aggregation process described in Section~\ref{ssec:latency} that combines more views of the same problem as the allowed latency increases.  Note that we kept the hyper-parameters ($\tau_{\mathrm{active}}$, $\rho_{\mathrm{update}}$, $\delta_{\mathrm{new}}$) optimized for latency $\lambda = 5$s and still get reasonable performance for lower latencies. However, it is also possible to re-optimize the hyper-parameters for a specific latency. This is what we did for the $\lambda=1$s  setting marked with $\bullet$ in Table~\ref{tab:results} for comparison with FlexSTB~\cite{stb2}. Not only do we get better overall performance, but our approach also has the advantage of a lower memory footprint, as it never ingests nor runs inference on more than 5s of audio at a time (compared to 100s of FlexSTB) and keeps a single vector per speaker in memory (compared to 100s of acoustic features and per-speaker scores in FlexSTB). Furthermore, our approach with $\lambda=3$s reaches the same performance as the official offline baseline~\cite{dihard3} of the DIHARD~III challenge ($25.5\%$ vs $25.4\%$). \\

\begin{figure}[htb]
    \centering
    \includegraphics[trim=0.0cm 0.5cm 0.0cm 0.25cm,clip,width=\linewidth]{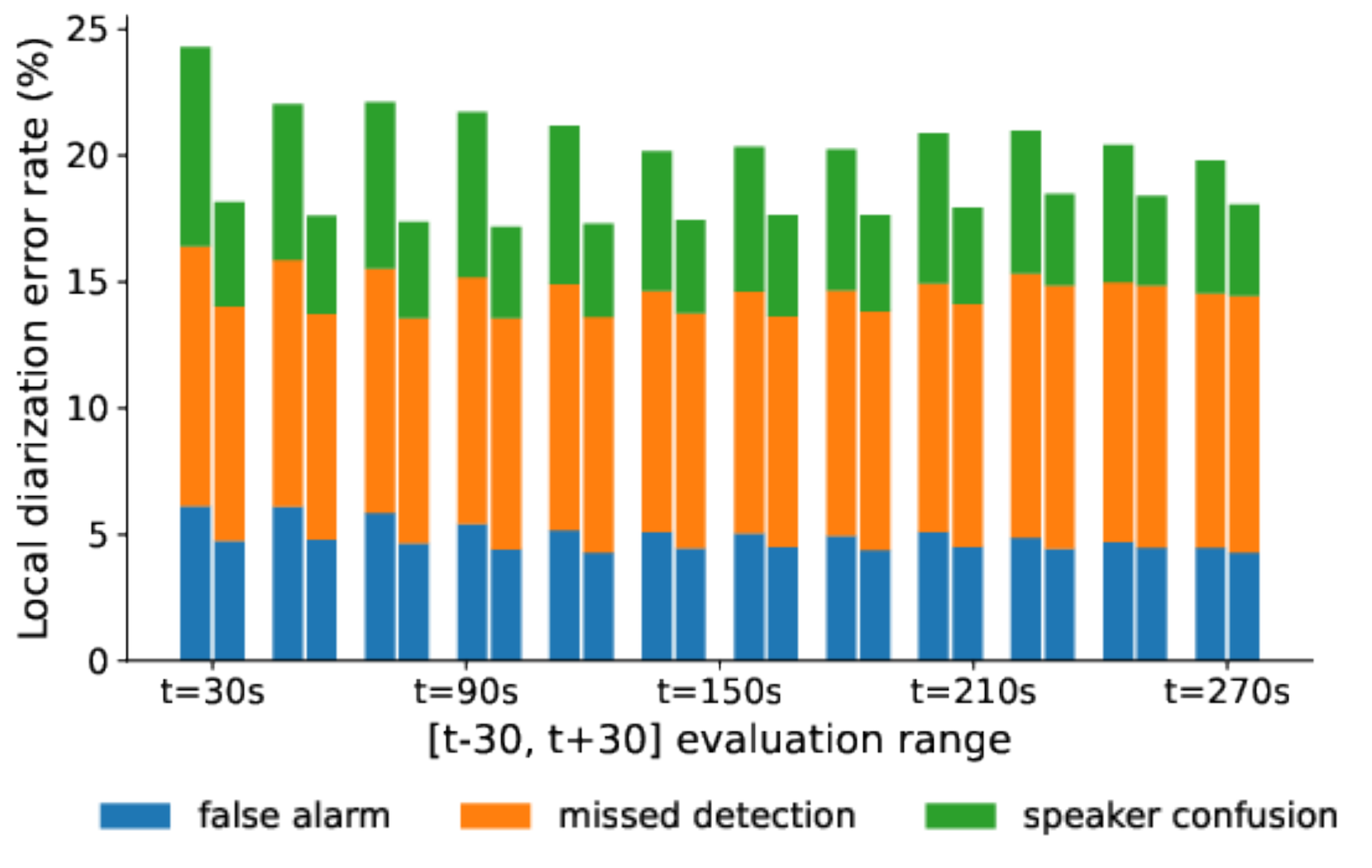}
    \caption{Evolution of performance as conversations unfold. Left: proposed online approach with 5s latency. Right: offline \textit{topline}~\cite{segmentation-herve}. Local diarization error rate computed on the 223 DIHARD~III (test) conversations that are longer than 300s.}
    \label{fig:der-evo}
    \vspace{-0.2cm}
\end{figure}

\vspace{-0.3cm}
\noindent\textbf{Continual learning}.
Figure~\ref{fig:der-evo} compares the performance over time of our online system against the offline \textit{topline}~\cite{segmentation-herve}.
While the performance of the latter remains somewhat constant, the former gets better as conversations unfold, almost bridging the gap after 5 minutes of conversation. As new information becomes available, our system learns better speaker centroids, hence decreasing speaker confusion error. 
While very long conversations can become rather expensive (if not impossible) to process with most offline models, our system can handle daylong audio streams at a practically constant memory cost, while getting better and better. \\

\vspace{-0.2cm}
\noindent\textbf{Reproducible research}. We share an open-source implementation of this work, as well as expected outputs in {\footnotesize{\texttt{RTTM}}} format at this address to facilitate future comparisons: {\footnotesize{\texttt{\href{https://github.com/juanmc2005/StreamingSpeakerDiarization/}{github.com/juanmc2005/StreamingSpeakerDiarization}}}}. \\

\vspace{-0.2cm}
\noindent\textbf{Real time}.
Computation time for one step of the rolling buffer is 165ms on a CPU Intel Cascade Lake 6248 (20 cores at 2.5Ghz) or 50ms on a GPU Nvidia Tesla V100 SXM2. This is suitable for real time applications, as the rolling buffer can be processed before its next update (every 500ms). 

\vspace{-0.3cm}
\section{Conclusion}
\label{conclusion}

\vspace{-0.1cm}
We have proposed an overlap-aware online speaker diarization system combining the end-to-end local segmentation of a 5 second long rolling buffer with incremental clustering.
Apart from handling overlapping speech at every stage, our system benefits from an adjustable latency between 500ms and 5s.
We show that our system outperforms FlexSTB~\cite{stb2} with a lower memory consumption, and that it is capable of bridging the gap to offline performance as conversations unfold.
This last advantage may make it preferable to an offline system when recordings are long and resources low.

\clearpage
\bibliographystyle{IEEEbib}
\bibliography{refs}

\end{document}